\documentclass[conference,10pt]{IEEEtran}
\usepackage{epsfig,epsf,rotating,setspace,latexsym,amsmath,amssymb,amsfonts,bm,theorem,subfigure,epstopdf,cite,authblk,bbm,nonfloat,comment}
\usepackage{algorithm}
\usepackage[noend]{algpseudocode}
\usepackage{color}
\usepackage{mathtools}
\usepackage{soul}

\usepackage[
top=0.75in,
bottom=1.1in,
left=0.64in,
right=0.65in]{geometry}

\algrenewcommand\algorithmicforall{\textbf{foreach}}
\algrenewcommand\algorithmicindent{.8em}

\newtheorem{lemma}{Lemma}

\newenvironment{Proof}[1]{\medskip\par\noindent{\bf Proof:\,}\,#1}{{\mbox{\,$\blacksquare$}\par}}

\IEEEoverridecommandlockouts
\allowdisplaybreaks
\begin{document}

\title{Age of Gossip in Ring Networks With \\ Non-Poisson Updates} 

\author{Arunabh Srivastava \qquad Sennur Ulukus\\
        \normalsize Department of Electrical and Computer Engineering\\
        \normalsize University of Maryland, College Park, MD 20742\\
        \normalsize  \emph{arunabh@umd.edu} \qquad \emph{ulukus@umd.edu}}

\maketitle

\begin{abstract}
    We consider a network consisting of $n$ nodes connected in a ring formation and a source that generates updates according to a renewal process and disseminates them to the ring network according to a Poisson process. The nodes in the network gossip with each other according to a push-based gossiping protocol, and disseminate version updates. Gossip between two neighbors happens at the arrivals of renewal processes with finite mean and variance. All renewal processes and Poisson processes in the network are independent but not identically distributed. We consider both uni-directional ring networks and bi-directional ring networks. We use version age of information to quantify the freshness of information at each node. Prior work has used the stochastic hybrid systems (SHS) approach or a first passage percolation (FPP) approach to analyze ring networks with edges following identical Poisson processes. In this work, we use a sample-path backtracking approach to characterize the probabilistic scaling of the version age of information of an arbitrary node in the gossip network, where each edge follows an independent but not identically distributed renewal process. We show that the version age of information of any node in the network is stochastically equivalent to $\sqrt{n}$ at any time instant after the node has received its first update from the source.
\end{abstract}

\section{Introduction}\label{sec: introduction}
Modern cyber-physical systems and large-scale internet of things (IoT) networks are becoming increasingly popular, creating the need for structured decentralized architectures. In many practical applications, such as perimeter surveillance arrays and vehicular networks, physical constraints determine specific spatial topologies, such as a ring or a line network. Due to their size and geographic isolation, these networks typically lack continuous centralized connectivity. Instead, these networks receive fresh updates by mobile or scheduled external sources such as uncrewed aerial vehicles (UAVs), mobile base stations, or even deterministic satellite links. Due to the physical trajectories or deterministic schedules of central sources, the update generation and dissemination process cannot be modeled using Poisson processes, which have been used traditionally in network analysis. Thus, there is a need to study information dissemination in ring and line networks under general non-Poisson renewal processes.

It is not possible to maintain network-wide information freshness by relying solely on intermittent direct updates from a band-limited central source. This limitation motivates the usage of localized peer-to-peer dissemination and gossip algorithms\cite{chettri2019comprehensive,swamy2020empirical}. In formations like ring networks, nodes communicate exclusively with their immediate neighbors. By forwarding updates locally, the network allows fresh information to propagate quickly throughout the network. This decentralized mechanism ensures that every node consistently receives fresh updates through its neighbors, even when the external central source, such as a UAV or a base station, is far away. This also enables scalable architecture, where a large number of nodes can cooperate successfully.

In such cellular and IoT networks, accurate real-time decision making heavily relies on information freshness. Usage of stale data can cause critical errors and inefficiencies in time-sensitive applications. To rigorously quantify information freshness, the age of information (AoI) metric has been widely adopted\cite{kaul2012real, sun2019age, yatesJSACsurvey}. Further needs to capture important network-dependent metrics led to the creation of other new metrics to measure freshness, including the age of incorrect information (AoII) \cite{maatouk20AOII}, the age of synchronization (AoS) \cite{zhong18AoSync}, and the binary freshness metric (BFM) \cite{cho3BinaryFreshness}. In systems where information updates represent discrete state changes, the version age of information \cite{yates21gossip, Abolhassani21version, melih2020infocom} serves as a precise metric.

In this work, we use the version age of information (simply, version age) to quantify the freshness of information. The version age of a node in a gossiping network is defined as the difference between the version at the source and the version at the node, where the source is generating updates and has the latest version. Ring networks following source updates were empirically studied for the first time in \cite{yates21gossip}. This work observed that the long-term average version age of nodes in the ring network is $\Theta(\sqrt{n})$. This observation was later proved in \cite{buyukates21CommunityStructure}. Later works examined extensions of the ring network, such as the generalized ring network \cite{srivastava2023generalizedrings}, and the line network \cite{kaswan22jamming}. Other adjacent works can be found in \cite{kaswan2025versionagesurvey, srivastava2023grid, mitra_allerton22, srivastava2024varyingtopologies}. Most works for this type of system model use Poisson processes on the edges to simplify analysis. \cite{kaswan25timeliness}, on the other hand, analyzed multi-hop networks, tree networks and fully-connected networks while considering edges having non-Poisson renewal processes. \cite{kaswan25timeliness} showed the effects of variance and path length on the long-term average version age of nodes in the network. However, except for the fully-connected network, there was a single path from the source to each node in all other models considered in the work. The fully-connected network was further analyzed using the inherent sparsity of the individual edge processes.

In this paper, we bridge the gap between structured ring topologies and general non-Poisson network dynamics. Unlike the prior studies that relied on the SHS or FPP framework, which strictly necessitated the Poisson process assumptions, we introduce a sample-path backtracking approach. We explicitly evaluate the transit times and inter-arrival times of updates over independent heterogeneous renewal processes with finite first and second moments on the edges of the ring network. For the uni-directional ring network, we perform a spatial window optimization to show that we only need to track updates coming to a node from a $\Theta(\sqrt{n})$ sized set of nodes that are directly upstream. Subsequently, we show that the version age of an arbitrary node in the uni-directional ring scales as $\Theta_p(\sqrt{n})$, where $\Theta_p(\cdot)$ denotes stochastic equivalence. We then extend the analysis to bi-directional ring networks. We show that the complexity introduced by the possibility of multiple paths existing to each node can be mitigated by a preemption argument. This argument states that long paths of size $\Theta_p(n)$ are preempted by their counterpart shorter paths of size $\Theta_p(\sqrt{n})$ with high probability. This result enables us to preserve the $\Theta_p(\sqrt{n})$ scaling for nodes in the bi-directional ring network that we observed previously in the uni-directional ring network.

The rest of the paper is organized as follows: In Section~\ref{sec: system model}, we define our system model. We then use the sample-path backtracking approach to find the average version age of each node in the uni-directional ring network in Section~\ref{sec: unidirectional ring}. We then show how this analysis can be extended for the bi-directional ring network in Section~\ref{sec: bidirectional ring}. We provide numerical simulations in Section~\ref{sec: numerical results} and conclude in Section~\ref{sec: conclusion}.

\section{System Model and the Version Age Metric}\label{sec: system model}
We consider a network consisting of a source, which is generating updates, and a gossip network consisting of $n$ nodes. The source generates updates on the arrivals of a renewal process with associated counting process $N_e$, and having inter-arrival times with finite mean $\mu_0 > 0$ and finite variance $\sigma^2_0$. The set of nodes in the gossip network is defined as $\mathcal{V} = \{0,1,\ldots,n-1\}$. The source sends updates directly to each node as a Poisson process with rate $\frac{\lambda_s}{n}$, thus sending updates to the entire gossip network with constant rate $\lambda_s$. 

\begin{figure}[t]
    \centering
    \includegraphics[width=0.63\linewidth]{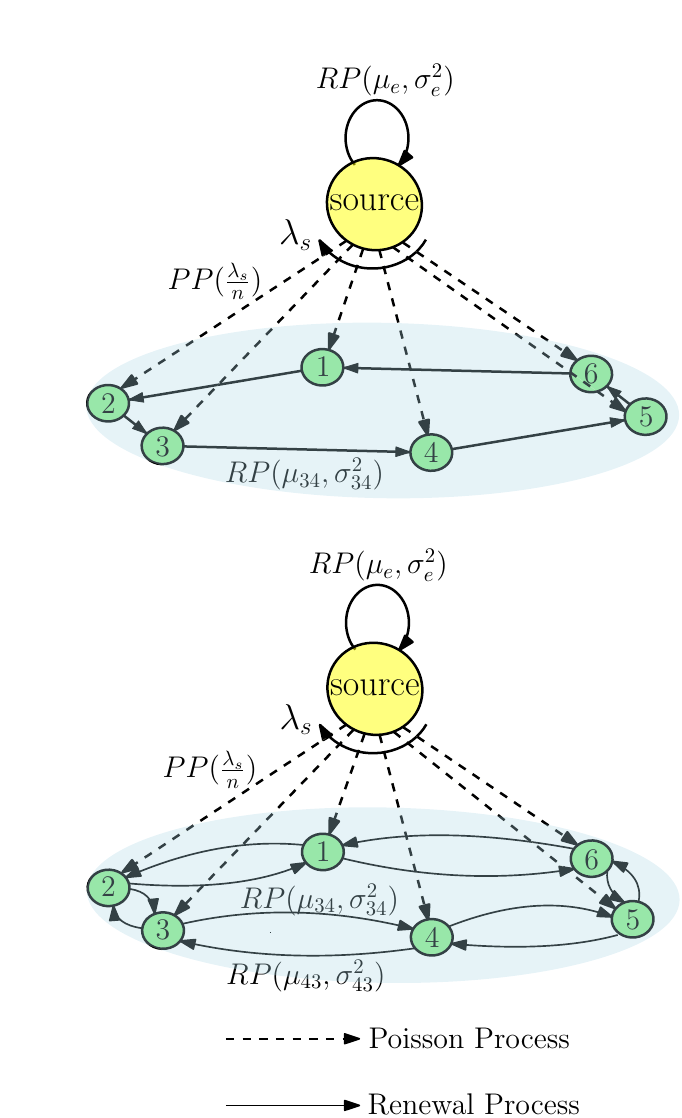}
    \caption{An illustration of the uni-directional ring network (top), and the bi-directional ring network (bottom). Note that all edges in the gossip network follow independent but not identically distributed renewal processes. The source generates updates according to a renewal process and sends updates to the gossip network as a Poisson process. Nodes gossip according to renewal processes which are not necessarily Poisson.}
    \label{fig: system model}
\end{figure}

We consider two types of ring networks, the uni-directional ring network and the bi-directional ring network. These are illustrated in Fig~\ref{fig: system model}. In the uni-directional ring network, node $\{i\}_n$ sends updates to node $\{i+1\}_n$ via the push gossiping protocol, where $\{i\}_n = i \text{ modulo }n$. In a similar way, in the bi-directional ring network, node $\{i\}_n$ sends updates to nodes $\{i-1\}_n$ and $\{i+1\}_n$ via the push gossiping protocol. Node $i$ sends updates to node $j$ at the arrival times of a renewal process that has inter-arrival times with mean $\mu_{ij} > 0$ and finite variance $\sigma_{ij}^2$. We assume that each parameter listed is a constant not dependent on $n$.

In order to quantify information freshness for the nodes in the network, we utilize the version age of information metric. Let $N_i(t)$ be the counting process associated with node $i$'s version update process. Then, the version age of node $i$ is given as $X_i(t) = N_e(t) - N_i(t)$. Under the push gossiping protocol, each node in the gossiping network sends updates to its neighbors. Each neighbor then chooses to keep the update if the new update's version is better than the version of the update it currently possesses. If this is not the case, then the neighboring node rejects the packet. Similarly, the source pushes updates to nodes in the network. Since the source always possesses the freshest update, every node accepts an update sent from the source, and in this case, the node's version age drops to $0$. Finally, if the source generates a new update, then every node's version age increases by $1$.

We assume that all renewal processes in the network are independent of each other. Further, we use the standard definitions for the stochastic boundedness $O_p$ and stochastic equivalence $\Theta_p$ of random variables. We also use the standard asymptotic notation for deterministic scalings $O,\Omega, \Theta, o, \omega$.
 
\section{Analysis of the Uni-Directional Ring}\label{sec: unidirectional ring}
In this section, we establish the scaling behavior of version age for nodes in the uni-directional ring topology. We select this structure as a foundational case for the analysis of networks with non-Poisson updates because its inherent structure eliminates complexities involving routing ambiguity and interference between different updates traveling throughout the network. In a uni-directional ring network, there is exactly one path from one node to another node. This allows us to isolate the impact of renewal processes on nodes' version age bypassing other underlying complexities. 

We first observe that the renewal processes governing the edges of the gossip network are heterogeneous. This results in varying version age statistics across different nodes. Therefore, without loss of generality, we focus our analysis on node $1$. In order to find a stochastic equivalent, i.e., $\Theta_p$, scaling for the version age of node $1$, we employ the sample path backtracking approach. In this approach, we pick an arbitrary update that was successfully accepted by node $1$, and retrace its total transit time from the source to node $1$. By evaluating the source counting process $N_e$ over this random transit time, we can directly quantify the number of new versions generated while the update was traveling. We note that as soon as the tracked update reaches node $1$, the node's version age drops to the exact number of times the source generated updates during the transit time of the update, which we shall show scales as $\Theta_p(\sqrt{n})$. This accounts for the scaling of the valleys in the version age process of node $1$.

Furthermore, we must account for the node's version age trajectory while the update was traveling or had not been generated yet. The version age of the node starts increasing when the successful update (the one that arrives at node $1$ before our tracked update) arrives. Thus, the version age of the node sees a peak right before our tracked update arrives. Therefore, rigorously bounding the peak version age requires us to evaluate the stochastic scaling of the update inter-arrival time at node $1$. To do so, we shall define a set of upstream nodes located within $\Theta(\sqrt{n})$ hops of node $1$. Using the Poisson process property of source to node updates, we shall show that the waiting time until this set of nodes receives an update directly from the source is $\Theta(\sqrt{n})$. Once a node in this set has the update, the update traverses at most $\Theta(\sqrt{n})$ edges to reach node $1$. We shall show that this time also scales as $\Theta_p(\sqrt{n})$. If a later update overtakes this one in the network, the inter-arrival time would only be shorter, which makes this bound a worst-case scenario. Thus, because both the transit time and inter-arrival time scale as $\Theta(\sqrt{n})$, we conclude that the peak version age is tightly bounded by the sum of both windows.

Finally, we note that no assumption is required about the starting distribution of the nodes' version age at $t=0$. Once the first update from the source arrives at a node, the previous history of version age becomes inconsequential. Therefore, we find that after the first successful update arrives at node $1$, the version age of node $1$ scales as the transit time of an arbitrary update, bounded tightly by $\Theta_p(\sqrt{n})$.

We have now established that the time scaling of both the transit time and the worst case inter-arrival time in the uni-directional ring dictates the scaling of the peak and valley of the version age process. We must now formally map these time durations to the actual version age metric. Since the source generates updates according to a renewal counting process $N_e$, we must evaluate the stochastic scaling of the random variable $N_e(T)$, where $T$ represents a random variable associated with travel times. To this end, we introduce the following lemma.

\begin{lemma}\label{lemma: 1}
    Suppose T is a random variable such that $\mathbb{E}[T] = \Theta(g(n))$. Then, $N_e(T) = O_p(g(n))$. Further, if $\text{Var}[T] = o((g(n))^2)$, then $N_e(T) = \Theta_p(g(n))$.
\end{lemma}

\begin{Proof}
    We use Markov's inequality to prove this result. First, we calculate the deterministic scaling for $\mathbb{E}[N_e(T)]$. We note the following,
    \begin{align}
        \frac{x}{\mu_e} - 1 < \mathbb{E}[N_e(x)] \leq \frac{x}{\mu_e}+\frac{\sigma_e^2+\mu_e^2}{\mu_e^2}.
    \end{align}
    The lower bound is a standard step in the proof of the \emph{elementary renewal theorem}, and the upper bound is due to Lorden's inequality \cite[Thm~1]{lorden1970excess}. This directly implies the following,
    \begin{align}
        \frac{T}{\mu_e} - 1 < \mathbb{E}[N_e(T)|T] \leq \frac{T}{\mu_e}+\frac{\sigma_e^2+\mu_e^2}{\mu_e^2}.
    \end{align}
    Taking expectations, we obtain,
    \begin{align}
        \frac{\mathbb{E}[T]}{\mu_e} - 1 < \mathbb{E}[N_e(T)] \leq \frac{\mathbb{E}[T]}{\mu_e}+\frac{\sigma_e^2+\mu_e^2}{\mu_e^2}.
    \end{align}
    This directly implies that $\mathbb{E}[N_e(T)] = \Theta(g(n))$. 
    
    Next, to show stochastic boundedness, given $\epsilon>0$, we find finite $M>0$ and finite integer $N_1>0$, such that,
    \begin{align}
        \mathbb{P}\bigg(\bigg|\frac{N_e(T)}{g(n)}\bigg|>M\bigg) < \epsilon, \quad \forall n \geq N_1.
    \end{align}
    We first choose $N_1$ such that $c_1 g(n) \leq \mathbb{E}[N_e(T)] \leq c_2 g(n)$, $\forall n \geq N_1$. Using Markov's inequality, we have,
    \begin{align}
        \mathbb{P}\bigg(\bigg|\frac{N_e(T)}{g(n)}\bigg|>M\bigg) \leq& \frac{\mathbb{E}[N_e(T)]}{Mg(n)}\\
        \leq& \frac{\lambda_e c_2}{M}.
    \end{align}
    Choosing $M > \frac{\lambda_e c_2}{\epsilon}$ completes the proof of the first part. 
    
    To prove the second part, we need to show that, given $\epsilon$,
    \begin{align}
        \mathbb{P}\big(N_e(T) > mg(n)\big) \geq 1 - \epsilon, \quad \forall n \geq N,
    \end{align}
    where $m>0$ is some positive constant, and $N>0$ is an integer. We first note that $\text{Var}[N_e(T)] = \mathbb{E}[\text{Var}[N_e(T)|T]]+\text{Var}[\mathbb{E}[N_e(T)|T]]$, due to the law of total variance. Due to the central limit theorem for renewal processes, we have that $\text{Var}[N_e(x)] \leq \frac{\sigma_e^2}{\mu_e^3}x + o(x)$. Therefore, $\text{Var}[N_e(T)|T] \leq \frac{\sigma_e^2}{\mu_e^3}T + o(T)$, which yields 
    \begin{align}
        \mathbb{E}[\text{Var}[N_e(T)|T]] \leq& \frac{\sigma_e^2}{\mu_e^3}\mathbb{E}[T] + o(\mathbb{E}[T])\\ =& O(g(n))\\
        =&o((g(n))^2).
    \end{align} 
    Further, for the second term, we establish from Lorden's inequality that $\mathbb{E}[N_e(T)|T] = {T}/{\mu_e}+e(T)$, where $-1 < e(T) \leq \frac{\sigma_e^2+\mu_e^2}{\mu_e^2}$. This implies that
    \begin{align}
        \text{Var}[\mathbb{E}[N_e(T)|T]] =& \text{Var}[T/\mu_e+e(T)]\\ 
        =& O(\text{Var}[T])\\
        =& o((g(n))^2),
    \end{align} 
    since $e(T)$, a bounded term, does not change the order of the variance. 
    Finally, to show stochastic equivalence, we apply Chebyshev's inequality. For a given $\delta>0$,
    \begin{align}
        \mathbb{P}\Big(\big|N_e(T) - \mathbb{E}[N_e(T)]\big| \geq \delta \mathbb{E}[N_e(T)]\Big) \leq \frac{\text{Var}[N_e(T)]}{\delta^2 (\mathbb{E}[N_e(T)])^2}.
    \end{align}
    We see that the numerator is $o((g(n))^2)$, and the denominator is strictly lower bounded by $\delta^2 (c_1 g(n))^2$, hence, the right hand side approaches $0$ as $n \rightarrow \infty$. Therefore, for any $\epsilon > 0$, there exists positive integer $N_2$ such that for all $n > \max(N_1,N_2)$, 
    \begin{align}
        \mathbb{P}\Big((1-\delta)\mathbb{E}[N_e(T)] < N_e(T) < (1+\delta)\mathbb{E}[N_e(T)]\Big) \geq 1 - \epsilon.
    \end{align}
    Since $\mathbb{E}[N_e(T)]=\Theta(g(n))$, $N_e(T)$ is bounded between positive constants multiplied with $g(n)$, with probability approaching $1$. Thus, we have $N_e(T)=\Theta_p(g(n))$, concluding the proof. 
\end{Proof}

We now evaluate the mean and variance of the inter-arrival time and transit time of an arbitrary update that arrives at node $1$. First, we focus on the inter-arrival time.

To evaluate the inter-arrival time between two successful updates at node $1$, we must analyze the time required for an update to be generated and sent to the network, and eventually propagate to node $1$. We note that the waiting time for an update to be sent to the network does not factor into the calculation of the transit time, as the transit time tracks the time strictly from the exact moment the update that eventually gets accepted by node $1$ was generated by the source. Howeer, the waiting time is critical for bounding the peak version age, as the node's version age continues to grow while the network waits for a successful update to arrive.

To rigorously prove that the peak version age scales as $\Theta_p(\sqrt{n})$, it is insufficient to show that some spatial window yields the correct inter-arrival time scaling. We must mathematically guarantee that no other spatial window could provide updates at a faster rate, thereby showing that $\Theta_p(\sqrt{n})$ is the minimum achievable scaling for the inter-arrival time. To this end, let $\mathcal{V}_k$ denote the set of $k$ consecutive upstream nodes immediately before node $1$ on the ring. Since the source updates each node independently according to a Poisson process with rate $\frac{\lambda_s}{n}$, $\mathcal{V}_k$ is updated by the source as a Poisson process with rate $\frac{k\lambda_s}{n}$. Therefore, the waiting time $W_k$ since the last update arrived at node $1$ until the source sends a new update to $\mathcal{V}_k$ is exponentially distributed with mean $\mathbb{E}[W_k] = \frac{n}{k\lambda_s}$, due to the memoryless property. Once the update arrives at $\mathcal{V}_k$, the update must traverse at most $k$ edges to reach node $1$. Let $T_{prop,k}$ denote this propagation time, and let $X_i$ denote the time spent at node $i$ along the path of $j$ nodes that has been realized. The path length is bounded by $k$, therefore, the expected propagation time is $\mathbb{E}[T_{prop,k}] = \sum_{i=1}^jX_i = O(k)$. The total inter-arrival time yielded by updates passing through this set is defined as $T_{inter,k} = W_k + T_{prop,k}$.

We observe that three regimes are possible in this case.
\begin{itemize}
    \item $k=o(\sqrt{n})$: In this regime, the nodes in $\mathcal{V}_k$ are very close to node $1$. This leads to $\mathbb{E}[T_{prop,k}] = o(\sqrt{n})$, which directly implies that the propagation time $T_{prop,k} = O_p(\sqrt{n})$ is very small. However, the source sends updates to $\mathcal{V}_k$ at a rate which is $o(\frac{1}{\sqrt{n}})$, since the set is very small. This leads to a high expected waiting time, $\mathbb{E}[W_k] = \omega(\sqrt{n})$, which directly implies that $W_k = \omega_p(\sqrt{n})$, and by extension, $T_{inter,k} = \omega_p(\sqrt{n})$. Therefore, while updates travel through the set very fast, the set itself does not receive updates fast enough to maintain $\Theta_p(\sqrt{n})$ inter-arrival time.
    \item $k = \omega(\sqrt{n})$: In this regime, the nodes in $\mathcal{V}_k$ can be very far from node $1$. The source sends updates to $\mathcal{V}_k$ at a rate which is $\omega(\frac{1}{\sqrt{n}})$, since the set is very large. This leads to a low expected waiting time, $\mathbb{E}[W_k] = o(\sqrt{n})$, which directly implies that $W_k = O_p(\sqrt{n})$. However, since the update has to travel through a large set of nodes, $\mathbb{E}[T_{prop,k}] = \omega(\sqrt{n})$, which directly implies that the propagation time $T_{prop,k} = \omega_p\sqrt{n}$ is very high. By extension, $T_{inter,k} = \omega_p(\sqrt{n})$. Therefore, while the set $\mathcal{V}_k$ receives updates from the source fast enough, the update is not able to propagate through the set fast enough.
    \item $k=\Theta(\sqrt{n})$: This regime achieves the optimal balance between the waiting time and propagation time. Since the source updates $\mathcal{V}_k$ with rate $\Theta(\frac{1}{\sqrt{n}})$, this implies that $\mathbb{E}[W_k] = \Theta(\sqrt{n})$, and since $W_k$ is exponentially distributed, $W_k = \Theta_p(\sqrt{n})$ and $N_e(W_k) = \Theta_p(\sqrt{n})$. Next, since the update needs to pass through $\Theta(\sqrt{n})$ nodes, we find that $\mathbb{E}[T_{prop,k}] = \Theta(\sqrt{n})$ and $\text{Var}[T_{prop,k}] = \Theta(\sqrt{n})$. Therefore, by applying Lemma~\ref{lemma: 1} to $T_{prop,k}$, we can conclude that $N_e(T_{prop,k}) = \Theta_p(\sqrt{n})$. Thus, we have that $N_e(T_{inter,k}) = \Theta_p(\sqrt{n})$.
\end{itemize}

After analyzing the above three regimes, we conclude that all updates that arrive at node $1$ arrive from a set of size $\Theta(\sqrt{n})$, and therefore, the inter-arrival time scales as $\Theta_p(\sqrt{n})$. Intuitively, the $\Theta(\sqrt{n})$ spatial window emerges as the strict minimizer of the version age due to the fundamental tradeoffs observed in the network dynamics. The version age is dictated by a balance between how often the source drops an update into a specific region, and the network propagation time for that region. If the region is too large, updates arrive into it frequently, but take too long to traverse the ring. If the region is too small, then the traversal time along the ring is small, but the source updates this region so infrequently that node $1$ is starved of fresh updates.

Next, we evaluate the transit time of an update that was successfully accepted by node $1$. In this case, the transit time is divided into two components: the waiting time until the update successfully reaches the ring, and the propagation time through the ring. Since the update reaches node $1$ successfully without being preempted, it was transmitted to the network before the next update was generated by the source. Since the source update generation process $N_0$ has a finite mean inter-generation time, the waiting time until the update successfully reaches the ring is $\Theta_p(1)$. The analysis of the propagation time follows through exactly as the spatial window optimization discussed in the analysis of the inter-arrival time, yielding that the propagation time is $\Theta_p(\sqrt{n})$. Therefore, the transit time of the update is the sum of these traversal times $\Theta_p(1)+\Theta_p(\sqrt{n})$, which scales as $\Theta_p(\sqrt{n})$. Further, the variance of the waiting time of the update is $\Theta(1) = o(n)$ and the variance of the propagation time is $\Theta(\sqrt{n}) = o(n)$. Therefore, due to the application of Lemma~\ref{lemma: 1}, the version age of node $1$ upon successfully receiving the update scales as $\Theta_p(\sqrt{n})$.

In conclusion, we found that the peak version age of node $1$ scales as the sum of the inter-arrival time and the transit time of an arbitrary update. Thus, the peak version age scales as $\Theta_p(\sqrt{n})$. Similarly, the valleys of the version age process of node $1$ scale as the transit time of successful updates that reach the node. Thus, the valleys scale as $\Theta_p(\sqrt{n})$ as well. This shows that, after node $1$ receives its first update, its version age scaling is $\Theta_p(\sqrt{n})$ for the rest of the time horizon in the uni-directional ring network. This also shows that the long-term average version age of any node in the network is $\Theta_p(\sqrt{n})$.

\section{Extension to the Bi-Directional Ring}\label{sec: bidirectional ring}
In the analysis of the bi-directional ring network, we encounter one additional layer of complexity when compared to the uni-directional ring network: path ambiguity. An update that was successfully accepted by node $1$ can traverse the ring in two opposite directions, creating a short path and a long path. In this section, we resolve this issue and show that for any node in the bi-directional ring network, the version age scaling remains $\Theta_p(\sqrt{n})$.

In order to prove this result, we re-deploy the sample-path backtracking approach and the spatial optimization technique we used for the uni-directional ring. We modify the arguments to account for the path ambiguity in the network. As discussed in Section~\ref{sec: unidirectional ring}, the peak version age is the sum of the inter-arrival time and transit time, while the valley of the version age is determined only by the transit time.

We first calculate the inter-arrival time. In order to account for the bi-directionality of the network, we redefine $\mathcal{V}_k$ to include $k$ consecutive nodes in the clockwise direction as well as the anti-clockwise direction, yielding a total of $2k$ nodes. Since the source updates nodes uniformly, $\mathcal{V}_k$ receives updates with rate $\frac{2k\lambda_s}{n}$. This introduces a constant factor change in the rate of $W_k$, and thus the scaling of $W_k$ and $N_e(W_k)$ remain identical to the uni-directional case.

Next, we look at the propagation time $T_{prop,k}$, and resolve the path ambiguity in the network. Suppose a node in $\mathcal{V}_k$ that is $d \leq k$ hops away from node $1$ receives the update from the source that eventually reaches node $1$. This update can reach node $1$ through two routes: the short path of $d$ hops, and the complementary long path of length $n-d$. When we evaluate the optimal spatial regime now, we must account for the longer path as well. We see that among the shorter paths, version age is minimized by paths that are $\Theta(\sqrt{n})$ in length. This means that the propagation time for the short path is $\Theta_p(\sqrt{n})$. The long path has length $n-\Theta(\sqrt{n}) = \Theta(n)$. Therefore, the propagation time along the long path is $\Theta_p(n)$. Since the short paths deliver the update to node $1$ in $\Theta_p(\sqrt{n})$ time, and the corresponding copy of the update would arrive along the longer path $\Theta_p(n)$ time later, node $1$ would receive the update through the shorter path. By the time the update arrives along the longer paths, it would be preempted by fresher updates with high probability. This implies that the long paths do not contribute to the version age process of node $1$. This allows us to bound the version age using the shorter paths.

Since we consider only the short paths in the bi-directional ring network, the bi-directional spatial window acts as two independent, overlapping spatial windows of the uni-directional network. Therefore, the three spatial regimes analyzed in Section~\ref{sec: unidirectional ring} remain unchanged:
\begin{itemize}
    \item If $k = o(\sqrt{n})$, the symmetric $\mathcal{V}_k$ is updated too infrequently, leading to $\omega_p(\sqrt{n})$ version age scaling.
    \item If $k = \omega(\sqrt{n})$, then the propagation time is too long, leading to $\omega_p(\sqrt{n})$ version age scaling.
    \item If $k = \Theta(\sqrt{n})$, the optimal balance is achieved again, and we are able to achieve $\Theta_p(\sqrt{n})$ version age scaling.
\end{itemize}

The same argument holds for the transit time of successful updates. The waiting time of the successful update at the source is $\Theta_p(1)$, and the short paths determine the optimal propagation time of $\Theta_p(\sqrt{n})$. We then use Lemma~\ref{lemma: 1} to conclude that the peaks, valleys and thus the long-term average version age of each node in the bi-directional network scale as $\Theta_p(\sqrt{n})$.

\section{Numerical Results}\label{sec: numerical results}

In this section, we perform numerical simulations for the bi-directional ring network where the source generates updates according to a renewal process with inter-renewal times distributed as $\text{Unif}[0,1]$. The source disseminates updates to the gossip network as a rate $1$ Poisson process. The number of nodes in the simulations increases from $100$ to $1000$ with a step size of $100$.

\begin{figure}[t]
    \centering
    \includegraphics[width=0.8\linewidth]{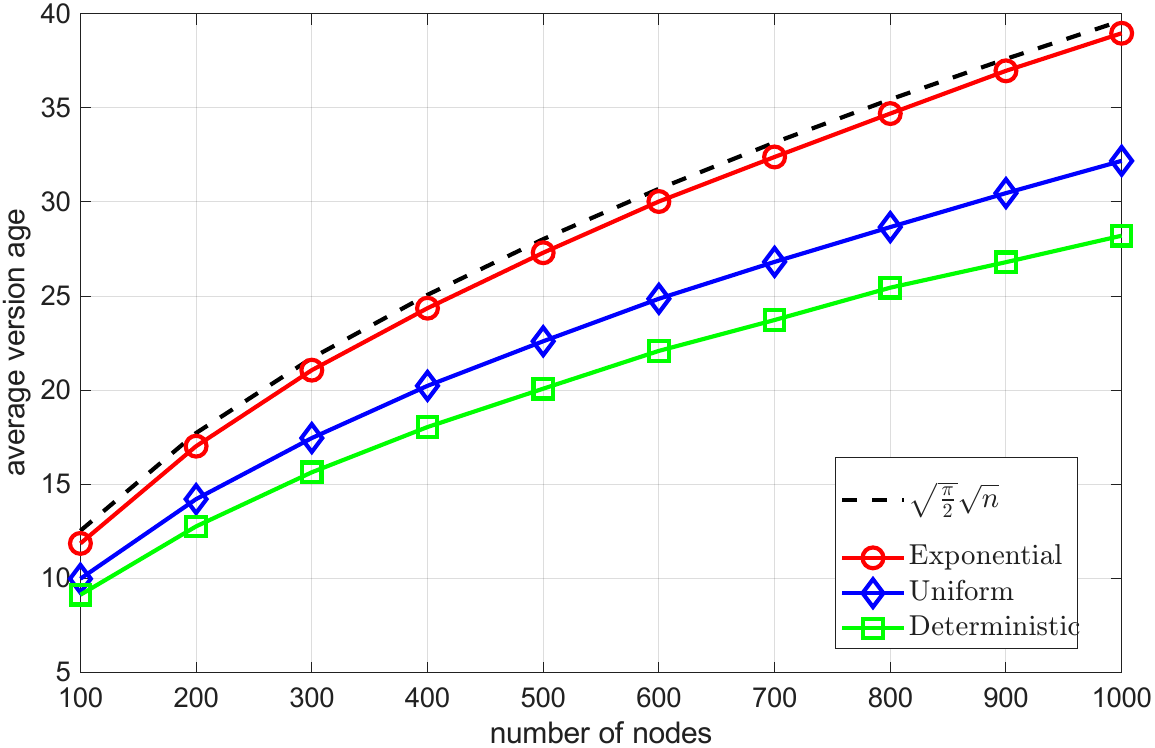}
    \caption{Version age of a network with i.i.d.~edge renewal processes.}
    \label{fig: iid plot}
\end{figure}

\begin{figure}[t]
    \centering
    \includegraphics[width=0.8\linewidth]{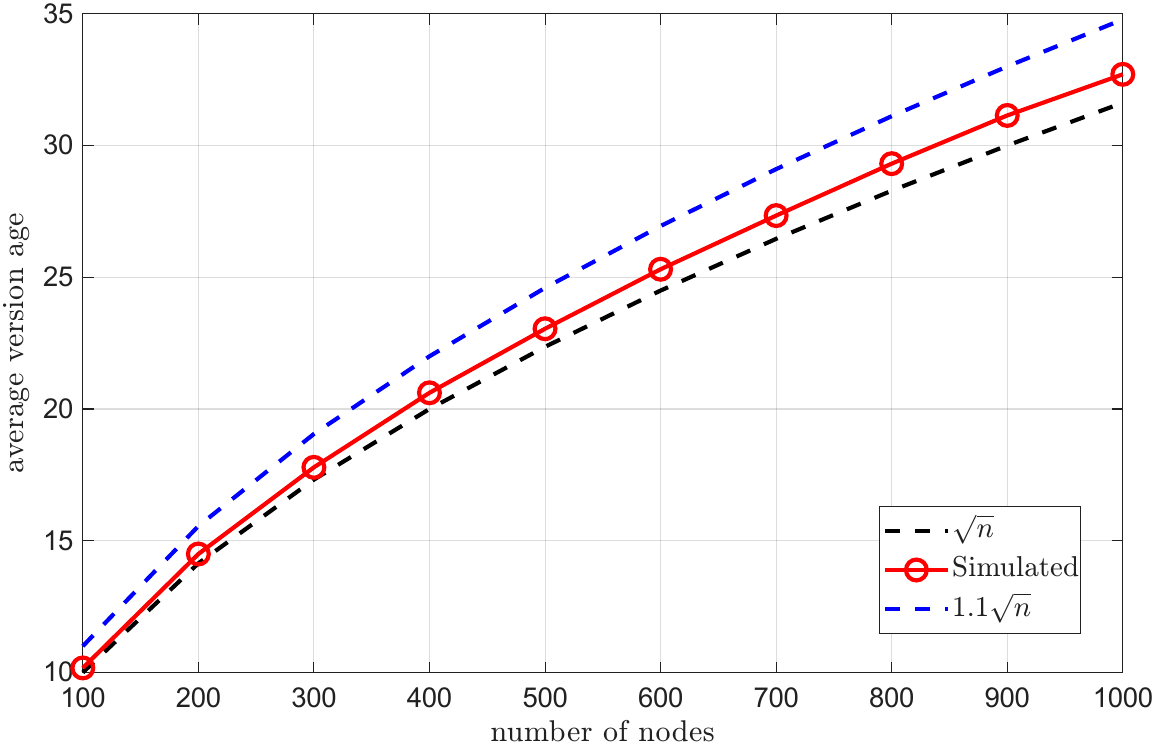}
    \caption{Version age of a network with non-i.i.d.~edge renewal processes.}
    \label{fig: non iid plot}
\end{figure}

\begin{figure}[t]
    \centering
    \includegraphics[width=0.8\linewidth]{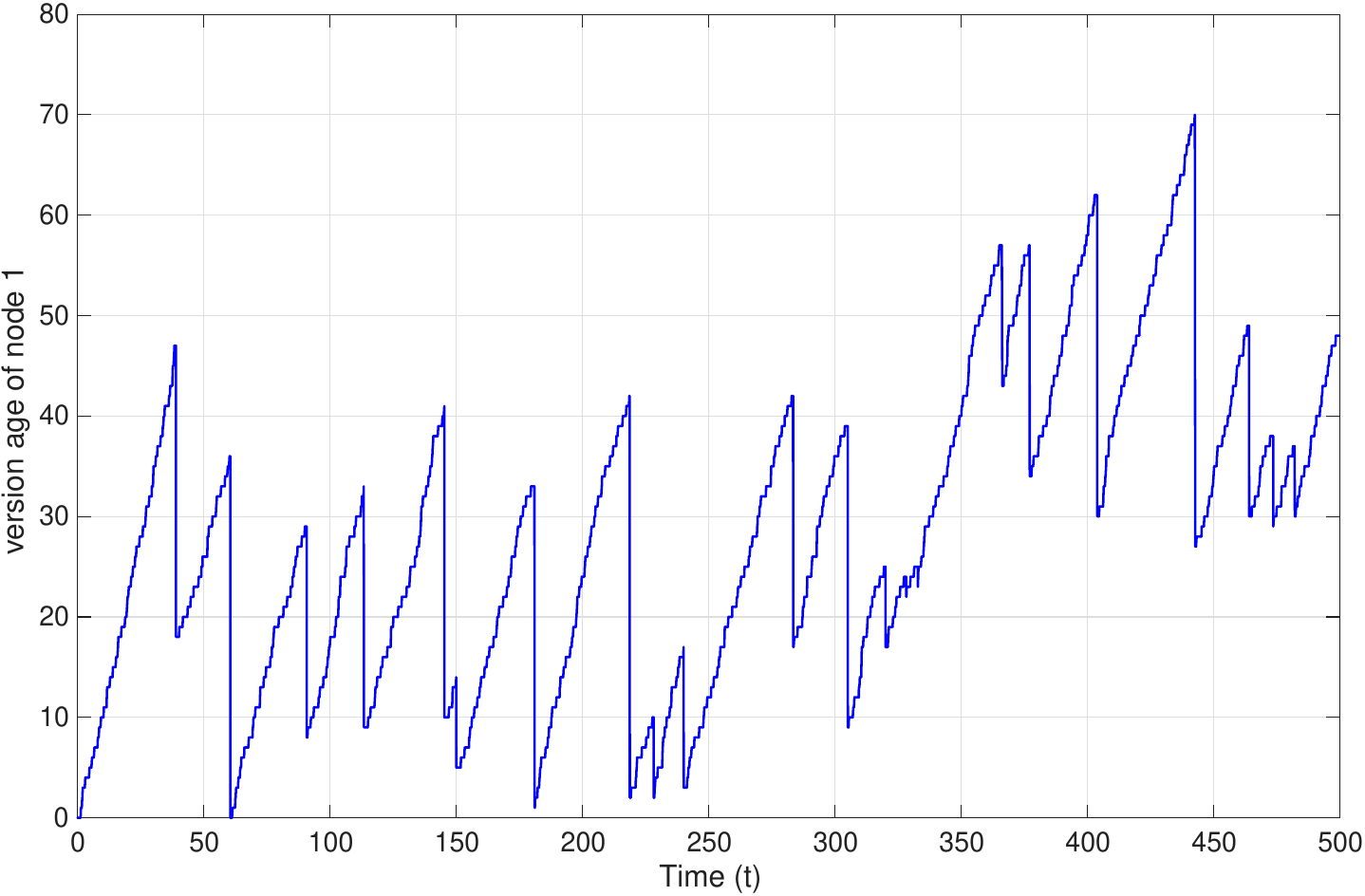}
    \caption{The evolution of version age of a single node.}
    \label{fig: single node version age evolution}
\end{figure}

In Fig.~\ref{fig: iid plot}, we consider the case where all renewal processes in the gossip network are i.i.d., and plot the average version age of the network as the number of nodes increases. We consider three inter-renewal distributions, the exponential distribution, which corresponds to Poisson processes, the uniform distribution, and the deterministic distribution. We choose a mean of $2$ in all cases. We observe that, as the variance of the inter-renewal distribution increases, the average version age of the network increases as well. Further, we also provide the theoretical bound of the long-term average version age of the bi-directional ring network found in \cite{buyukates21CommunityStructure} for comparison. This simulation supports our result in Section~\ref{sec: bidirectional ring} that the version age of the nodes scales as $\Theta_p(\sqrt{n})$.

Next, in Fig.~\ref{fig: non iid plot}, we consider the case where all renewal processes are independent but not identically distributed. We assign each edge one of the following distributions as its inter-renewal time randomly: exponential distribution, uniform distribution, deterministic distribution, Gamma distribution with shape parameter $2$, Rayleigh distribution, $\chi$-square distribution with $4$ degrees of freedom and Beta distribution using shape parameters $\alpha=2$ and $\beta=2$. We scale each distribution so that its mean is $2$. Using the $\Theta(\sqrt{n})$ upper bound and lower bound, we show that the average version age of the network behaves as predicted by Section~\ref{sec: bidirectional ring}. In Fig.~\ref{fig: single node version age evolution}, we plot how the version age of node $1$ evolves in this network for the first $500$ time units. This node was a part of a bi-directional ring network with a $1000$ nodes. We observe that the node sees fresher packets very regularly at $O(\sqrt{n})$ intervals as predicted in Section~\ref{sec: bidirectional ring}, even though they might have high version age themselves, as is evident from the later part of the plot.

\section{Conclusion}\label{sec: conclusion}
In this paper, we characterized the scaling laws of information freshness in uni-directional and bi-directional ring networks under general renewal processes. Using a sample path backtracking approach, we proved that the version age in both types of ring networks, as well as in finite line networks, scale as $\Theta_p(\sqrt{n})$. Our simulations show that this asymptotic behavior is robust to popular underlying delay distribution, while the Poisson processes serve as an upper bound for the version age among these renewal processes due to their high variance. Furthermore, our methodology establishes identical scaling for the age of information metric as well.

\bibliographystyle{unsrt}
\bibliography{refs}

\begin{thebibliography}{10}

\bibitem{chettri2019comprehensive}
L.~Chettri and R.~Bera.
\newblock A comprehensive survey on internet of things ({I}o{T}) toward {5G} wireless systems.
\newblock {\em IEEE Internet of Things Journal}, 7(1):16--32, October 2019.

\bibitem{swamy2020empirical}
S.~N. Swamy and S.~R. Kota.
\newblock An empirical study on system level aspects of internet of things ({I}o{T}).
\newblock {\em IEEE Access}, 8:188082--188134, October 2020.

\bibitem{kaul2012real}
S.~K. Kaul, R.~D. Yates, and M.~Gruteser.
\newblock Real-time status: How often should one update?
\newblock In {\em IEEE Infocom}, March 2012.

\bibitem{sun2019age}
Y.~Sun, I.~Kadota, R.~Talak, and E.~Modiano.
\newblock Age of information: A new metric for information freshness.
\newblock {\em Synthesis Lectures on Communication Networks}, 12(2):1--224, December 2019.

\bibitem{yatesJSACsurvey}
R.~D. Yates, Y.~Sun, D.~R. Brown, S.~K. Kaul, E.~Modiano, and S.~Ulukus.
\newblock Age of information: An introduction and survey.
\newblock {\em IEEE Jour. on Selected Areas in Communications}, 39(5):1183--1210, May 2021.

\bibitem{maatouk20AOII}
A.~Maatouk, S.~Kriouile, M.~Assaad, and A.~Ephremides.
\newblock The age of incorrect information: A new performance metric for status updates.
\newblock {\em IEEE/ACM Trans. on Networking}, 28(5):2215--2228, October 2020.

\bibitem{zhong18AoSync}
J.~Zhong, R.~D. Yates, and E.~Soljanin.
\newblock Two freshness metrics for local cache refresh.
\newblock In {\em IEEE ISIT}, June 2018.

\bibitem{cho3BinaryFreshness}
J.~Cho and H.~Garcia-Molina.
\newblock Effective page refresh policies for web crawlers.
\newblock {\em ACM Trans. on Database Systems}, 28(4):390--426, December 2003.

\bibitem{yates21gossip}
R.~D. Yates.
\newblock The age of gossip in networks.
\newblock In {\em IEEE ISIT}, July 2021.

\bibitem{Abolhassani21version}
B.~Abolhassani, J.~Tadrous, A.~Eryilmaz, and E.~Yeh.
\newblock Fresh caching for dynamic content.
\newblock In {\em IEEE Infocom}, May 2021.

\bibitem{melih2020infocom}
M.~Bastopcu and S.~Ulukus.
\newblock Who should {G}oogle {S}cholar update more often?
\newblock In {\em IEEE Infocom}, July 2020.

\bibitem{buyukates21CommunityStructure}
B.~Buyukates, M.~Bastopcu, and S.~Ulukus.
\newblock Age of gossip in networks with community structure.
\newblock In {\em IEEE SPAWC}, September 2021.

\bibitem{srivastava2023generalizedrings}
A.~Srivastava and S.~Ulukus.
\newblock Age of gossip on generalized rings.
\newblock In {\em IEEE MILCOM}, October 2023.

\bibitem{kaswan22jamming}
P.~Kaswan and S.~Ulukus.
\newblock Age of gossip in ring networks in the presence of jamming attacks.
\newblock In {\em Asilomar Conference}, October 2022.

\bibitem{kaswan2025versionagesurvey}
P.~Kaswan, P.~Mitra, A.~Srivastava, and S.~Ulukus.
\newblock Age of information in gossip networks: A friendly introduction and literature survey.
\newblock {\em IEEE Transactions on Communications}, 73(8):6200--6220, February 2025.

\bibitem{srivastava2023grid}
A.~Srivastava and S.~Ulukus.
\newblock Age of gossip on a grid.
\newblock In {\em Allerton Conference}, September 2023.

\bibitem{mitra_allerton22}
P.~Mitra and S.~Ulukus.
\newblock {ASUMAN}: Age sense updating multiple access in networks.
\newblock In {\em Allerton Conference}, September 2022.

\bibitem{srivastava2024varyingtopologies}
A.~Srivastava, T.~J. Maranzatto, and S.~Ulukus.
\newblock Age of gossip with time-varying topologies.
\newblock In {\em IEEE ISIT}, June 2025.

\bibitem{kaswan25timeliness}
P.~Kaswan and S.~Ulukus.
\newblock Timeliness in cache-aided networks with non-poisson updating.
\newblock {\em IEEE Transactions on Communications}, 73(8):6068--6080, August 2025.

\bibitem{lorden1970excess}
G.~Lorden.
\newblock On excess over the boundary.
\newblock {\em The Annals of Mathematical Statistics}, 41(2):520--527, April 1970.

\end{thebibliography}

\end{document}